\documentclass[aps,prl,letterpaper,twocolumn,showpacs,preprintnumbers,amsmath,amssymb,superscriptaddress]{revtex4}

\usepackage{graphicx}
\usepackage{dcolumn}
\usepackage{bm}
\usepackage{hyperref}

\usepackage{color}

\usepackage{CJK}

\begin{document}

\begin{CJK*}{UTF8}{gbsn}

\preprint{draft version: \today}


\title{Measurement of the Hyperfine Quenching Rate of the Clock Transition in $^{171}$Yb}

\author{C.-Y. Xu (徐晨昱)}
\affiliation{Physics Division, Argonne National Laboratory, Argonne, Illinois 60439, USA}
\affiliation{Department of Physics and Enrico Fermi Institute, University of Chicago, Chicago, Illinois 60637, USA}
\author{J. Singh}
\altaffiliation{Present address: Technische Universit\"{a}t M\"{u}nchen, Exzellenzcluster Universe, 85748 Garching, Germany.}
\affiliation{Physics Division, Argonne National Laboratory, Argonne, Illinois 60439, USA}
\author{J. C. Zappala}
\affiliation{Physics Division, Argonne National Laboratory, Argonne, Illinois 60439, USA}
\affiliation{Department of Physics and Enrico Fermi Institute, University of Chicago, Chicago, Illinois 60637, USA}
\author{K. G. Bailey}
\affiliation{Physics Division, Argonne National Laboratory, Argonne, Illinois 60439, USA}
\author{M. R. Dietrich}
\affiliation{Physics Division, Argonne National Laboratory, Argonne, Illinois 60439, USA}
\author{J. P. Greene}
\affiliation{Physics Division, Argonne National Laboratory, Argonne, Illinois 60439, USA}
\author{W. Jiang (蒋蔚)}
\affiliation{Physics Division, Argonne National Laboratory, Argonne, Illinois 60439, USA}
\author{N. D. Lemke}
\affiliation{Physics Division, Argonne National Laboratory, Argonne, Illinois 60439, USA}
\author{Z.-T. Lu (卢征天)}
\affiliation{Physics Division, Argonne National Laboratory, Argonne, Illinois 60439, USA}
\affiliation{Department of Physics and Enrico Fermi Institute, University of Chicago, Chicago, Illinois 60637, USA}
\author{P. Mueller}
\affiliation{Physics Division, Argonne National Laboratory, Argonne, Illinois 60439, USA}
\author{T. P. O'Connor}
\affiliation{Physics Division, Argonne National Laboratory, Argonne, Illinois 60439, USA}



\date{\today}

\begin{abstract}
We report the first experimental determination of the hyperfine quenching rate of the $6s^2\ ^1\!S_0\ (F=1/2) - 6s6p\ ^3\!P_0\ (F=1/2)$ transition in $^{171}$Yb with nuclear spin $I=1/2$.
This rate determines the natural linewidth and the Rabi frequency of the clock transition of a Yb optical frequency standard.
Our technique involves spectrally resolved fluorescence decay measurements of the lowest lying $^3\!P_{0,1}$ levels of neutral Yb atoms embedded in a solid Ne matrix.
The solid Ne provides a simple way to trap a large number of atoms as well as an efficient mechanism for populating $^3\!P_0$.
The decay rates in solid Ne are modified by medium effects including the index-of-refraction dependence.
We find the $^3\!P_0$ hyperfine quenching rate to be $(4.42\pm0.35)\times10^{-2}\ \mathrm{s}^{-1}$ for free $^{171}$Yb, which agrees with recent {\it ab initio} calculations.
\end{abstract}


\maketitle

\end{CJK*}

The conservation of angular momentum strictly forbids single-photon transitions between two atomic states if both electronic angular momenta are equal to zero, i.e. $J=0\nleftrightarrow J'=0$.
This restriction can be circumvented by state mixing due to the hyperfine interaction \cite{bowen36}.
The consequent increase in the transition rate is referred to as hyperfine quenching (HFQ), a feeble mechanism that typically plays a significant role in the radiative decay of only the lowest lying $^3\!P_{0,2}$ levels of divalent atoms.

The earliest studies of the HFQ effect focused on the spectra originating from nebulae \cite{garstang62}.
More recently, the isotopic dependence of these astronomical spectra have been used to infer HFQ rates \cite{bjp02} and, conversely, isotope ratios that result from stellar nucleosynthesis \cite{rfch04}.
In the laboratory, the $1s2p\ ^3\!P_{0,2}$ levels in He-like ions were the first to be measured and are the most thoroughly studied \cite{johnson11}.
The HFQ rates of a handful of many-electron ions have also been measured \cite{johnson11,cu+,s+,in+}.
However, the rate has never been measured in any neutral atoms due to difficulties involved in populating the relevant levels and subsequently observing their slow decay.

In neutral atoms, efforts have been made in modern \textit{ab initio} calculations of the HFQ rate \cite{pd04,scg04}, motivated by the promising application of neutral divalent atoms to optical clocks \cite{dk11}, quantum computing \cite{daley11}, and quantum simulation of many-body systems \cite{bdz08}.
In the case of optical clocks, the HFQ rate determines the natural linewidth and the Rabi frequency of the ``clock transition'' $ns^2\ ^1\!S_0 - {nsnp\ ^3\!P_0}$ in fermionic isotopes.
The HFQ rate calculations require accurate knowledge of the atomic structure of the many-electron atoms.
For $6s^2\ ^1\!S_0\ (F=1/2)-6s6p\ ^3\!P_0\ (F=1/2)$ in $^{171}$Yb with nuclear spin $I=1/2$, the HFQ rate ($A_\mathrm{HFQ}$) involves the matrix element of the electric-dipole operator ($D$) between intermediate levels ($\gamma$) and the ground level $6s^2\ ^1\!S_0$, as well as the hyperfine interaction ($H_\mathrm{HFI}$) matrix element between these levels ($\gamma$) and $6s6p\ ^3\!P_0$,
\begin{equation}
A_\mathrm{HFQ}(^1\!S_0-{^3\!P_0})\!\propto\!\left|\sum_\gamma\frac{\left<^1\!S_0||D||\gamma\right>\!\left<\gamma||H_\mathrm{HFI}||^3\!P_0\right>}{E(\gamma)-E(^3\!P_0)}\right|^2\!,
\label{eq:A-HFQ}
\end{equation}
where $E(\gamma)$ are the energies of atomic levels relative to the ground level \cite{pd04}.
Among the intermediate levels, the HFQ of $6s6p\ ^3\!P_0$ is predominantly caused by the admixture of the lowest lying $6s6p\ ^3\!P_1$ and $6s6p\ ^1\!P_1$ \cite{pd04}, from which the transitions to the ground level are both E1 allowed.
Our measurement, therefore, serves as a sensitive benchmark for these calculations.

\begin{figure}[b]
\centering
\includegraphics[width=0.35\textwidth]{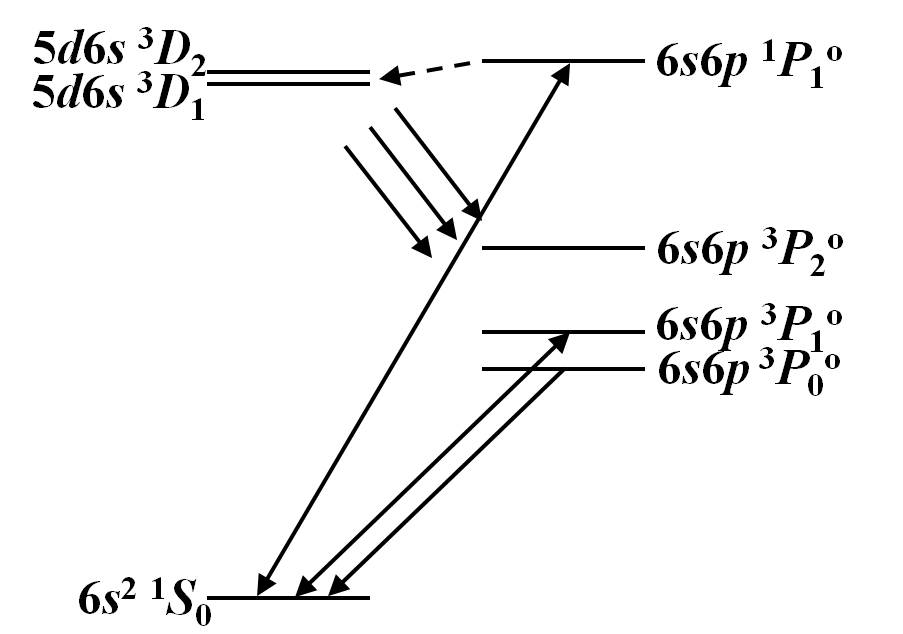}
\caption{
Low-lying atomic levels and transitions of Yb in solid Ne.
$^3\!P_0$ can be efficiently populated by virtue of an enhanced inter-system crossing ${^3\!D_1} \leftarrow {^1\!P_1}$.
The radiative decay of $^3\!P_0$ is observed in both $^{171}$Yb and $^{172}$Yb samples.
\label{fig:Yb}}
\end{figure}

We employ a novel technique of probing atoms embedded in solid Ne to extract the HFQ rate $A_\mathrm{HFQ}(^1\!S_0-{^3\!P_0})$ in free $^{171}$Yb.
Interrogating atoms trapped in a solid offers both high atomic density and long observation time.
In addition, while matrix isolated Yb atoms qualitatively resemble free atoms, they exhibit an enhanced inter-system crossing $5d6s\ ^3\!D_1 \leftarrow 6s6p\ ^1\!P_1$, enabling efficient population of $6s6p\ ^3\!P_0$ by pumping the strong $6s^2\ ^1\!S_0 - 6s6p\ ^1\!P_1$ transition and subsequent spontaneous decay (Fig.~\ref{fig:Yb}) \cite{xu11}.
We choose solid Ne as the matrix because it is less polarizable than heavier noble gas solids and more technically accessible than solid He.
While He only solidifies under at least 25 bar pressure, Ne readily forms a solid with face-centered-cubic crystal structure at 24.5 K and 1 bar \cite{ashcroft}.

The main challenge of performing this measurement is to properly account for various medium effects.
First, the medium may open additional radiative or non-radiative decay channels on an excited atom.
Second, the medium may alter the HFQ rate of a free atom by modifying the atomic wavefunctions and shifting the energies in Eq.~(\ref{eq:A-HFQ}).
Third, Fermi's golden rule dictates that the spontaneous emission rate of a transition depends cubically on the transition frequency that may be shifted in medium.
Finally, the spontaneous emission rate also depends on the environment of the emitter.
Such a phenomenon, known as the Purcell effect, is one of the hallmarks of quantum electrodynamics (QED).
In cavity QED, the decay rate is modified by the geometry of the surrounding vacuum environment \cite{cqed,drexhage70,hctf87,nv11}.
Within a medium, however, the decay rate depends on the index of refraction because it modifies both the photon dispersion relation and the energy fluctuation of the QED vacuum.
Although the index-of-refraction effect has been known for some time, there is still considerable tension in its understanding \cite{toptygin03,db12}.

\begin{figure}[b]
\centering
\includegraphics[width=0.48\textwidth]{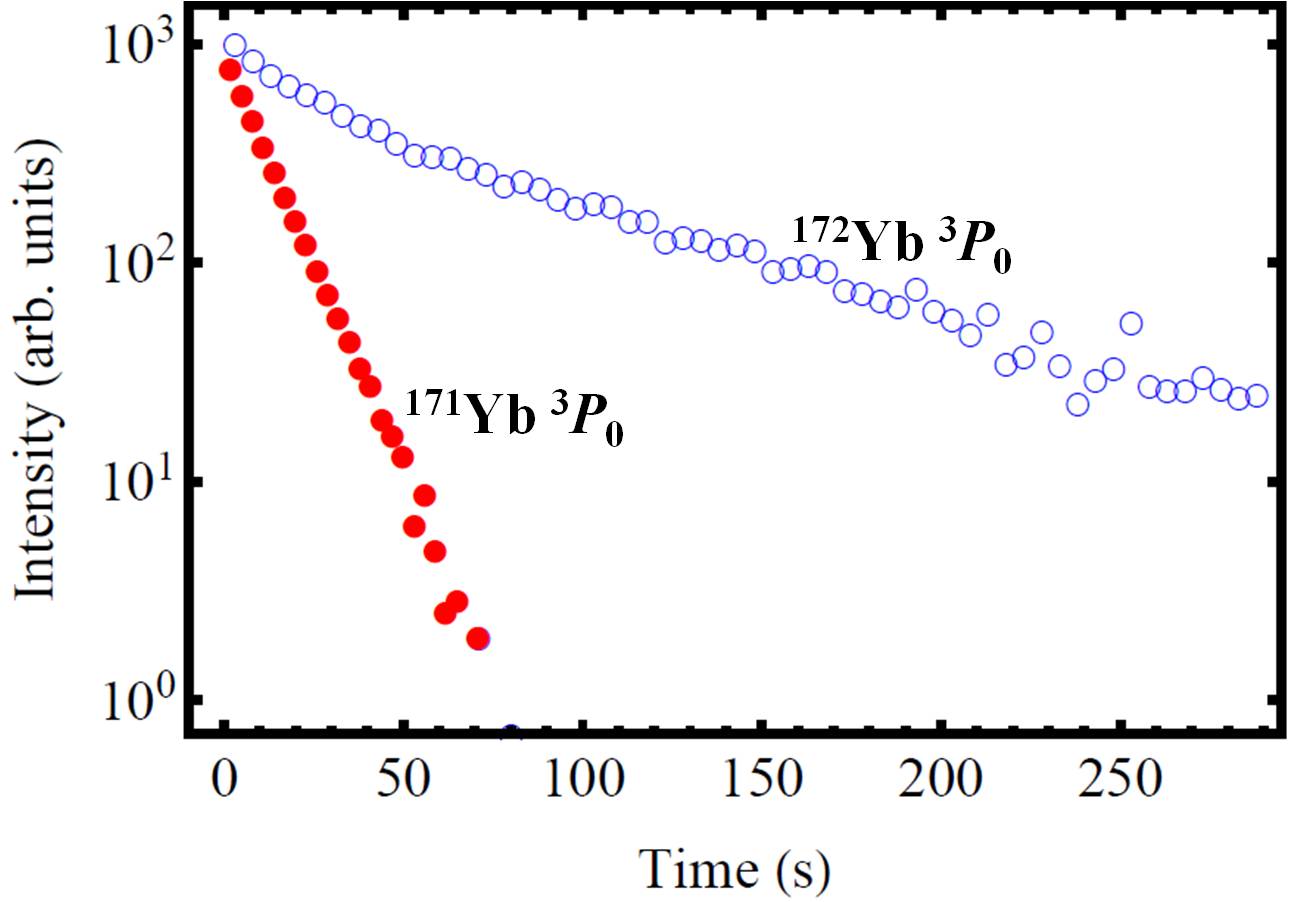}
\caption{
The time-dependent fluorescence intensity of $^{171}$Yb $^3\!P_0$ (red solid circles) and $^{172}$Yb $^3\!P_0$ (blue open circles) in solid Ne near the center of the emission peak.
The influence of the HFQ effect is evident.
\label{fig:3P0-decay}}
\end{figure}

We address these effects as follows.
First, we measure the $6s6p\ ^3\!P_0$ decay rate for isotopically pure $^{171}$Yb and $^{172}$Yb ($I=0$) in solid Ne (Fig.~\ref{fig:3P0-decay}).
The difference between these two rates separates the HFQ contribution from any medium quenching mechanisms that are independent of isotopes.
Second, the Yb transition frequencies in solid Ne are used to calculate the energy and frequency dependent corrections.
Third, we measure the decay rate of Yb $6s6p\ ^3\!P_1$ in solid Ne and compare it with the experimental value in vacuum \cite{beloy12} to provide a direct calibration of the index-of-refraction effect.
After making these corrections, we then obtain the HFQ rate of a free atom.

The samples are prepared with a similar setup we used previously \cite{xu11}.
Before the deposition on the liquid-He cooled sapphire substrate, Ne gas (99.999~\%) flows through a noble gas purifier (LDetek LDP1000) and a 77 K charcoal trap in order to minimize the growth defects and increase the sample transparency.
We co-deposit Yb using an atomic beam generated by an effusive oven.
To avoid the formation of Yb clusters, we keep the Yb-to-Ne ratio below 5 ppm and the temperature below 5 K to suppress the mobility of the atoms.
Samples with isotopically pure $^{171}$Yb (95~\%, Oak Ridge batch 196043) and $^{172}$Yb (97~\%, Oak Ridge batch 124501) are separately made.
While the enriched Yb is available for several even isotopes, $^{172}$Yb contains the least concentration of odd isotopes.

\begin{figure}[b]
\centering
\includegraphics[width=0.48\textwidth]{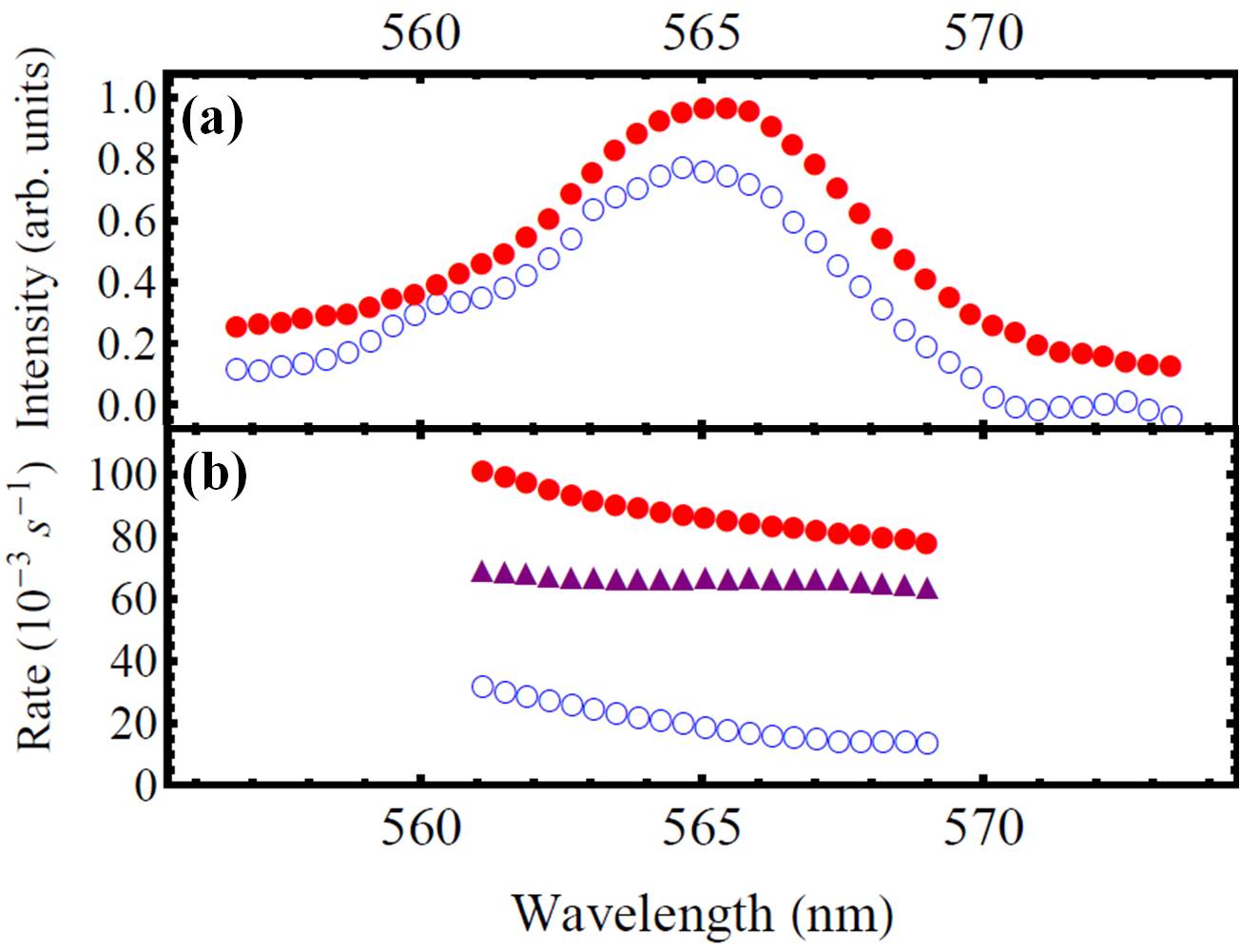}
\caption{
(a) The $^1\!S_0 \leftarrow {^3\!P_0}$ emission spectrum of $^{171}$Yb (red solid circles) and $^{172}$Yb (blue open circles) in solid Ne after the 385 nm LED is switched off.
The peak is shifted from the vacuum position at 578.4 nm.
The spectra are separately normalized so that the peaks appear to have similar height.
(b) The decay rate of $^{171}$Yb $^3\!P_0$ (red solid circles) and $^{172}$Yb $^3\!P_0$ (blue open circles), and the difference of the decay rates between the two isotopes (purple solid triangles) at select wavelengths.
The error bars are about the size of the markers.
\label{fig:3P0}}
\end{figure}

We use a 385 nm light-emitting diode (LED) to excite the $^1\!S_0 - {^1\!P_1}$ transition \cite{lambo12} and subsequently populate $^3\!P_0$.
The fluorescence is detected by a 1.5 nm resolution optical spectrometer (Ocean Optics USB4000-UV-VIS).
Fig.~\ref{fig:3P0}(a) shows the emission spectrum of $^{171}$Yb $^3\!P_0$ (solid circles) and $^{172}$Yb $^3\!P_0$ (open circles) in solid Ne after the LED is switched off.
We record the fluorescence decay at select wavelengths for 100 s for $^{171}$Yb and 300 s for $^{172}$Yb.
The decay near the center of the emission peak (565 nm) is shown in Fig.~\ref{fig:3P0-decay}.
At each wavelength, the decay rate of each isotope is obtained by fitting the data to an exponential function and is plotted in Fig.~\ref{fig:3P0}(b).
The uncertainty about the size of the markers is determined from the fitting error and the sample variance based on three $^{171}$Yb and two $^{172}$Yb samples with different Yb densities and optical transparencies.
The strong wavelength dependence of the decay rates is in part due to the frequency cube dependence.
The remainder is likely caused by the interaction with phonons of different energies.
The sum of multiple exponential functions, which describes multiple types of trapping sites in solid Ne, gives a better fit at some wavelengths, but the weighted average of the multiple rates is not significantly different from the rate of the single exponential fit.

The decay rate of $^{172}$Yb $^3\!P_0$ near 565 nm is approximately $2\times10^{-2}\ \mathrm{s}^{-1}$.
Since its single-photon decay in vacuum is strictly forbidden, this rate reflects the overall medium quenching of an excited atom.
One possible quenching mechanism may be that the atomic wavefunctions are perturbed by the crystal field in solid Ne.
To model this perturbation, we assume that this field is randomly oriented and has a constant strength.
Eleven low-lying levels are included, between which the reduced matrix elements of the electric-dipole operator have been calculated \cite{prk99}.
We sum over $M_J$ states and the orientation of the field, leading to a Stark-like coupling between levels.
In order to account for the observed decay rate, it requires a 20 MV/m crystal field so that the perturbed $^3\!P_0$ wavefunction has an admixture of $^3\!P_1$ with a mixing coefficient of $1.3\times10^{-4}$ and of $^1\!P_1$ with $1.7\times10^{-6}$.
Such a crystal field strength is not unexpected in solid Ne \cite{ashcroft}.

For the $^{171}$Yb $^3\!P_0$ decay, since the nuclear spins are unpolarized and the crystal field is randomly oriented, the effects of the medium quenching and the HFQ add incoherently.
We plot the difference of the decay rates between the two isotopes in Fig.~\ref{fig:3P0}(b) (solid triangles).
As expected, this differential rate is mostly independent of the wavelength and represents the HFQ contribution.
We take the average of the rates weighted by the emission intensity and find the HFQ rate of $^{171}$Yb $^3\!P_0$ in solid Ne to be $(6.72\pm0.28)\times10^{-2}\ \mathrm{s}^{-1}$.
The uncertainty is conservatively chosen to be half of the full range.

We first examine the medium's influence on the HFQ mechanism described in Eq.~(\ref{eq:A-HFQ}).
From the crystal field strength estimation, we are assured that the atomic wavefunctions are essentially intact.
However, the medium alters the energy differences in the denominators.
The HFQ of $^3\!P_0$ is predominantly caused by the admixture of the lowest lying $^3\!P_1$ and $^1\!P_1$ \cite{pd04}.
In solid Ne, we take $E(^3\!P_0)=(565\ \mathrm{nm})^{-1}$, $E(^3\!P_1)=(546\ \mathrm{nm})^{-1}$, and $E(^1\!P_1)=(396\ \mathrm{nm})^{-1}$ in the emission mode \cite{xu11,lambo12}.
$E(^3\!P_1)-E(^3\!P_0)$ in solid Ne is equal to $616\ \mathrm{cm}^{-1}$ and is changed from its vacuum value ($704\ \mathrm{cm}^{-1}$) by a factor of 0.875.
Therefore, the HFQ rate is enhanced by a factor of 1.306 if the $^3\!P_1$ term dominates the sum.
Similarly, $E(^1\!P_1)-E(^3\!P_0)$ is changed by 0.971, and the rate enhanced by 1.061.
Assuming a uniform probability distribution of the relative contribution from $^3\!P_1$ and $^1\!P_1$, we take the mid-point as the mean and $1/\sqrt{12}$ of the full range as the uncertainty \cite{jcgm100} and obtain an enhancement factor of $1.183\pm0.071$.

We then consider the medium's influence on the spontaneous emission rate ($A$) of a transition,
\begin{equation}
\frac{A_m}{A_v} = \left ( \frac{\omega_m}{\omega_v} \right)^3 G(n),
\label{eq:A-ratio}
\end{equation}
where $\omega$ is the transition frequency, the subscript $m$ ($v$) refers to medium (vacuum), and the scale factor $G$ is a function of the index of refraction ($n$).
To extract $A_{\mathrm{HFQ},v}(^1\!S_0-{^3\!P_0})$, we use $\omega_\mathrm{Ne}(^1\!S_0-{^3\!P_0})=(565\pm1\ \mathrm{nm})^{-1}$ and $\omega_v(^1\!S_0-{^3\!P_0})=(578.4\ \mathrm{nm})^{-1}$ to calculate the frequency dependent correction.
The uncertainty is due to the spectrometer calibration and the sample variance.

\begin{figure}[b]
\centering
\includegraphics[width=0.48\textwidth]{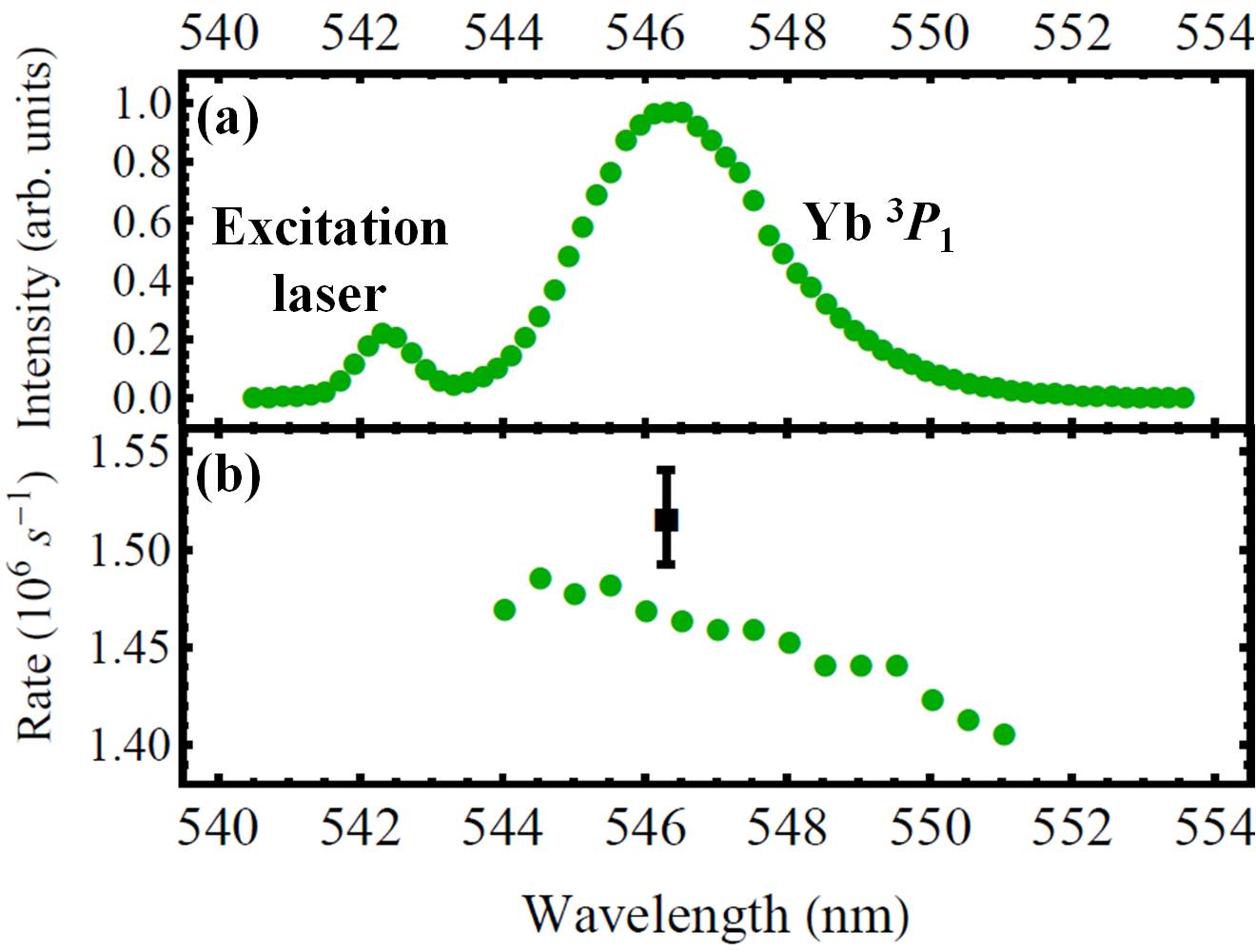}
\caption{
(a) The $^1\!S_0 \leftarrow {^3\!P_1}$ emission spectrum of Yb in solid Ne induced by the 543 nm laser.
The peak is shifted from the vacuum position at 555.8 nm.
(b) The decay rate of $^3\!P_1$ at select wavelengths in solid Ne (green circles).
The error bars are about the size of the markers.
The decay rate in vacuum, $1.162\times10^6\ \mathrm{s}^{-1}$, is off the scale.
The black square with an error bar indicates the predicted $^3\!P_1$ decay rate in solid Ne using the RC model and the frequency cube dependence.
\label{fig:3P1}}
\end{figure}

We determine $G_\mathrm{Ne}$ by measuring the $^3\!P_1$ decay for the following reasons.
The HFQ transition $^1\!S_0\ (F=1/2) - {^3\!P_0}\ (F=1/2)$ and the intercombination transition $^1\!S_0 -\ ^3\!P_1$ are both of E1 type.
Their transition wavelengths are sufficiently close that the wavelength dependence of the index of refraction is insignificant.
The $^3\!P_1$ decay rate in vacuum is precisely known $A_v(^1\!S_0-{^3\!P_1})=(1.162\pm0.008)\times10^6\ \mathrm{s}^{-1}$ \cite{beloy12}.
Compared to this rate, the medium quenching rate ($\sim2\times10^{-2}\ \mathrm{s}^{-1}$) is negligible, which allows us to use the measured total decay rate for $A_\mathrm{Ne}(^1\!S_0-{^3\!P_1})$.

For the $^3\!P_1$ lifetime measurement in solid Ne, samples of natural Yb are used.
We excite the $^1S_0 - {^3\!P_1}$ transition by a 543 nm diode pumped solid state laser (Opto Engine MGL-III-543).
The fluorescence light is coupled into a monochromator (McPherson 225) and detected by a photomultiplier tube counting module (Sens-Tech P10PC-2) mounted at the exit of the monochromator.
A dead time correction is applied for the counting rate.
Fig.~\ref{fig:3P1}(a) shows the steady-state emission spectrum of $^3\!P_1$ in solid Ne with 1 nm resolution.

We chop the laser at 50 kHz with 50~\% duty cycle using an acousto-optic modulator and record the decay at select wavelengths with 50 ns resolution.
The decay rate at each wavelength is plotted in Fig.~\ref{fig:3P1}(b).
The average of the rates weighted by the emission intensity is $A_\mathrm{Ne}(^1\!S_0-{^3\!P_1})=(1.464\pm0.040)\times10^6\ \mathrm{s}^{-1}$, where the uncertainty is half of the full range.
From Eq.~(\ref{eq:A-ratio}) for the $^3\!P_1$ decay with $\omega_\mathrm{Ne}(^1\!S_0-{^3\!P_1})=(546\pm1\ \mathrm{nm})^{-1}$ and $\omega_v(^1\!S_0-{^3\!P_1})=(555.8\ \mathrm{nm})^{-1}$, we obtain the transition-independent $G_\mathrm{Ne}=1.194\pm0.036$.
Again using Eq.~(\ref{eq:A-ratio}) for the $^3\!P_0$ decay, we arrive at $A_{\mathrm{HFQ},v}(^1\!S_0-{^3\!P_0})=(4.42\pm0.35)\times10^{-2}\ \mathrm{s}^{-1}$ for free $^{171}$Yb.
All the corrections we have made are summarized in Table \ref{tab:summary}.

We compare this result to two available calculations: $6.2\times10^{-2}\ \mathrm{s}^{-1}$ (no uncertainty provided) \cite{mb01} and $4.35\times10^{-2}\ \mathrm{s}^{-1}$ (a few percent uncertainty) \cite{pd04}.
Authors of reference \cite{mb01} have used experimentally measured hyperfine parameters in their calculation and have included only two intermediate levels in Eq.~(\ref{eq:A-HFQ}).
Authors of reference \cite{pd04} have computed the sum with multiple intermediate levels and have independently calculated the hyperfine constants with better than 1~\% accuracy as a verification of the quality of their technique.
Our measurement is in good agreement with reference \cite{pd04}.

\begin{table}[b]
\caption{
A summary of corrections due to medium effects for the extraction of the $^3\!P_0$ HFQ rate of free $^{171}$Yb based on the measurements in solid Ne.
\label{tab:summary}}
\centering
\begin{tabular}{l c c}
\hline \hline
Correction & Scale factor & Uncertainty \\
\hline
Energy difference, Eq.(\ref{eq:A-HFQ}) & 0.845 & 6.0~\% \\
Medium quenching & 0.771 & 4.2~\% \\
Index-of-refraction effect, Eq.(\ref{eq:A-ratio}) & 0.838 & 3.0~\% \\
Frequency cube, Eq.(\ref{eq:A-ratio}) & 0.932 & 0.5~\% \\
\hline
Total & 0.508 & 7.9~\%\\
\hline \hline
\end{tabular}
\end{table}

We are also able to compare our experimentally determined $G_\mathrm{Ne}$ to theoretical predictions.
One theory supported by recent experiments for E1 transitions \cite{rikken95,schuurmans98} is the real cavity (RC) model \cite{gl91}.
It treats the emitter as residing in an empty spherical cavity carved out of a lossless, homogeneous, and isotropic medium with permittivity $\epsilon=n^2\epsilon_0$.
The macroscopic field in the dielectric is canonically quantized.
The model predicts the following scaling with $n$,
\begin{equation}
G^\mathrm{RC}(n)=n^3\left[\frac{1}{n}\left(\frac{E_\mathrm{loc}}{E_\mathrm{mac}}\right)_\mathrm{RC}\right]^2.
\label{FRC}
\end{equation}
The factor $n^3$ comes from the in-medium photon dispersion relation.
The macroscopic field operator $\hat{\bf{E}}_\mathrm{mac}$ is renormalized by $1/n$ due to the in-medium energy density $\epsilon\hat{\bf{E}}_\mathrm{mac}^2/2$.
The ratio of the local field inside the cavity $E_\mathrm{loc}$ to the macroscopic field far outside the cavity $E_\mathrm{mac}$ is found to be $(E_\mathrm{loc}/E_\mathrm{mac})_\mathrm{RC}=3\epsilon/(2\epsilon+\epsilon_0)$ using the boundary conditions on the sphere.

Given the growth conditions of our solid Ne samples and both the wavelength- and the temperature- dependence of the index of refraction, we take $n_\mathrm{Ne}=1.10\pm0.01$ \cite{bls67,dell03,sk74}.
Therefore, the RC model gives $G^\mathrm{RC}_\mathrm{Ne}=1.239\pm0.024$.
This is in good agreement with our experimentally determined value $G_\mathrm{Ne}=1.194\pm0.036$.
The predicted $^3\!P_1$ rate in solid Ne using the RC model and the frequency cube dependence is also indicated in Fig.~\ref{fig:3P1}(b) (solid square).

In heavier noble gas solids, we find that the Yb transitions suffer from exacerbated medium effects.
In solid Ar, they manifest in a stronger wavelength dependence of the $^3\!P_1$ decay rate.
Our measurements show $\omega_\mathrm{Ar}(^1\!S_0-{^3\!P_1})=(562\pm1\ \mathrm{nm})^{-1}$, $A_\mathrm{Ar}(^1\!S_0-{^3\!P_1})=(1.82\pm0.19)\times10^{6}\ \mathrm{s}^{-1}$, and thus $G_\mathrm{Ar}=1.62\pm0.17$.
The larger uncertainty makes solid Ar a less attractive medium for transition-rate measurements.
Nevertheless, this result still agrees with the RC model prediction $G^\mathrm{RC}_\mathrm{Ar}=1.69\pm0.05$ with $n_\mathrm{Ar}=1.28\pm0.02$ \cite{ss69,sk74}.
In solid Xe, the Yb $^3\!P_0$ lifetime is shorter than 50 $\mu$s due to a much stronger crystal field.
Therefore, the HFQ measurement becomes impossible.
 
In conclusion, we have measured the HFQ rate of the $^1\!S_0 (F=1/2) - {^3\!P_0 (F=1/2)}$ transition in $^{171}$Yb based on the matrix isolation technique using solid Ne and spectrally resolved fluorescence decay measurements.
We have accounted for medium effects using measurements of both the $^{172}$Yb $^3\!P_0$ decay and the Yb $^3\!P_1$ decay in solid Ne.
The average $^3\!P_1$ decay rate across the emission peak in solid Ne agrees with the RC model prediction.
In order to carry out a more precise study on the index-of-refraction effect, one needs to consider the phonon interaction to better understand the wavelength dependence.

Finally, the most suitable naturally abundant candidates for the study of the HFQ effect using this technique are $^{25}$Mg, $^{43}$Ca, $^{67}$Zn, $^{87}$Sr, $^{111}$Cd, $^{113}$Cd, $^{171}$Yb, $^{173}$Yb, $^{199}$Hg, and $^{201}$Hg.
For each of these candidates, a naturally abundant nuclear spin-0 isotope is available, and the transition from the ground level to the lowest lying $^1\!P_1$ is optically accessible.
Lighter atoms are more tightly bound which likely means that the medium induced corrections are smaller but the efficiency of populating $^3\!P_0$ is worse.
On the other hand, lighter atoms also have higher $^1\!P_1$ levels which may provide an alternative and more efficient path for the $^3\!P_0$ population.

We would like to thank T. Oka, S. T. Pratt and R. W. Dunford for helpful discussions and the use of their equipment.
This work is supported by Department of Energy, Office of Nuclear Physics, under Contract No. DEAC02-06CH11357.
J. S. and N. D. L. are supported by Argonne Director's postdoctoral fellowship.



\end{document}